\documentclass{jfm}
\usepackage{graphicx}
\usepackage{color}

\shorttitle{Effects of polymer additives in the bulk of turbulent thermal convection}
\shortauthor{Y.-C. Xie, S.-D. Huang, D. Funfschilling, X.-M. Li, R. Ni and K.-Q. Xia}

\title{Effects of polymer additives in the bulk of turbulent thermal convection}

\author{Yi-Chao Xie$^1$, Shi-Di Huang$^1$, Denis Funfschilling$^2$, Xiao-Ming Li$^1$, Rui Ni$^1$\footnotemark[2] and Ke-Qing Xia$^1$\footnotemark[1]
\footnotetext[1]{Email address for correspondence: kxia@phy.cuhk.edu.hk}
\footnotetext[2]{Present address: Department of Mechanical and Nuclear Engineering, Pennsylvania State University, State College, PA 16802-1412, USA.}}
\affiliation{$^1$Department of Physics, The Chinese University of Hong Kong, Shatin, Hong Kong, China\\
$^2$LRGP, Lorraine University, CNRS, 1 rue Grandville B.P. 20451, F-54001 Nancy, France}

\begin{document}

\maketitle

\begin{abstract}
We present experimental evidence that a minute amount of polymer additives can significantly enhance heat transport in the bulk region of turbulent thermal convection. The effects of polymer additives are found to be the \textit{suppression} of turbulent background fluctuations that give rise to incoherent heat fluxes that make no net contribution to heat transport, and at the same time to \textit{increase} the coherency of temperature and velocity fields. The suppression of small-scale turbulent fluctuations leads to more coherent thermal plumes that result in the heat transport enhancement. The fact that polymer additives can increase the coherency of thermal plumes is supported by the measurements of a number of local quantities, such as the extracted plume amplitude and width, the velocity autocorrelation functions and the velocity-temperature cross-correlation coefficient. The results from local measurements also suggest the existence of a threshold value for the polymer concentration, only above which can significant modification of the plume coherent properties and enhancement of the local heat flux be observed. Estimation of the plume emission rate suggests that the second effect of polymer additives is to stabilize the
thermal boundary layers.
\end{abstract}


\section{Introduction}\label{sec:Intro}
	The ability to efficiently transport heat, momentum and mass is one of the hallmarks of turbulent flows, which finds numerous examples in both industry and daily life. One way to enhance mass transport, or equivalently to reduce wall-drag, is to seed the turbulent flow with tiny amount of long-chain polymers \citep*{toms1948some}. The so-called polymer-induced drag-reduction has been studied extensively \citep*[see, for example,][]{Procaccia2008RMP}. Compared to drag-reduction, the effects of polymer additives on turbulent heat transport, via turbulent thermal convection, for instance, has just started drawing attention very recently \citep*{Ahlers2010PRL,Benzi2010PRL,Boffetta2010PRL,Wei2012PRE}.  

	An idealized model to investigate turbulent thermal convection is the turbulent Rayleigh-B\'enard convection (RBC), which is a closed system of a fluid layer confined between two horizontally parallel plates heated from below and cooled from above \citep*{ahlers2009RMP,Xia2010ARFM,SchumacherChilla2012,Xia2013TAML}. This system is controlled by two dimensionless parameters, i.e. the Rayleigh number $Ra=\alpha g \Delta T H^3/(\nu\kappa)$ and the Prandtl number $Pr=\nu/\kappa$, where $g$ is the gravitational acceleration constant, $\Delta T$ the temperature difference across the plates separated by a distance $H$, $\alpha$, $\nu$ and $\kappa$ are, respectively, the thermal expansion coefficient, kinematic viscosity and thermal diffusivity of the fluid. The salient feature of turbulent RBC is the existence of distinct flow regions, i.e. the laminar, albeit fluctuating, thermal boundary layers (BLs) in the vicinity of top and bottom walls \citep*{Sun2008JFM,Zhou2010PRL,Zhou2010JFM} and a turbulent bulk region far away from the walls. The flow in the bulk region has been shown to be close to homogeneous turbulent thermal convection \citep*{2003PRL_Lohse,2005PoF_Cal}.

	With the presence of polymers, one thus needs to consider these two regions separately owing to their different flow dynamics. It has been shown in turbulent bulk flows in which temperature is not a dynamic variable that the energy cascade process is strongly modified by polymer additives \citep*{2008NJP_Eberhard,2007_JFM_Ouelltte,2013_PRL_Xi}. In the bulk flow of turbulent RBC where there are strong temperature and velocity fluctuations, it is proposed that polymer additives can enhance the heat transport \citep*{Benzi2010PRL}. In contrast, it is suggested that polymer additives stabilize the thermal BL, which reduces the emission of thermal plumes and hence the heat transport \citep{Ahlers2010PRL,Wei2012PRE}. The global Nusselt number $Nu$ (the total heat flux normalized by the one that would prevail by conduction alone) is then determined by the competition of these two effects. Although the aforementioned ideas could provide a plausible explanation to the observed heat transport enhancement in homogeouse turbulent thermal convection and heat transport reduction in the boundary-layer-dissipation dominated regime of turbulent RBC, direct evidence based on local measurements is still missing, which motivates the present study of the effects of polymer additives in the bulk of turbulent RBC.

\section{The experiments}\label{sec:exp}

	The experiment was carried out in a cylindrical convection cell with diameter $D$ (height $H$) being $19.3$ cm ($19.5$ cm).  Deionized and degassed water was used as the working fluid. Detailed design of the convection cell has been reported elsewhere \citep*{Qiu2005JoT,Wei2012PRE}. Here we mention only the essential features. The convection cell was an upright plexiglass cylinder with aluminium rough top and bottom plates whose surfaces were coated with a thin layer of Teflon. The roughness was in the form of pyramids with a height of 8 mm that were arranged in a square lattice, and was machined directly on the plates. The vertex angle of the roughness was $90^o$. It is suggested that in the present parameter range of $Ra$ and $Pr$ both energy and thermal dissipations are bulk-dominated in convection cells with rough surfaces compared to those in the cases with smooth surfaces \citep{Wei2014JFM}, which motivates us to carry out experiments in the rough cell.

	The convective heat flux, i.e. local Nusselt number, at the cell centre was measured using a small thermistor (Measurement Specialties, G22K7MCD419) combined with a two dimensional laser Doppler velocimetery (LDV, Dantec Dynamics), which is similar to those used in \citet{Shang2003PRL, Shang2004PRE}. The thermistor used has a time constant of 30 ms in water and a diameter of 0.3 mm. The spatial separation between the LDV focusing point and the thermistor tip was $\sim$0.8 mm, which is much smaller than the correlation length between temperature and velocity fluctuations \citep{Shang2004PRE}. The temperature and velocity measurements were synchronized in such a way that when a velocity burst was captured by the LDV processor, it would record the temperature simultaneously. Using the acquired temperature $T(t)$ and velocity $u(t)$ with an averaged sampling rate of $\sim$60 Hz, we obtain the instantaneous normalized local convective heat flux,	
	\begin{equation}
	 J_{\alpha}(t)=(u_{\alpha}(t)\times\delta T(t))~H/(\kappa\Delta T),
	\end{equation} 
where $\alpha$=$z$ and $x$ for the vertical and horizontal directions respectively; $\delta T(t)=T(t)-T_0$ is the temperature deviation from the mean temperature $T_0$ measured at cell centre. In addition, the temperature signal itself was separately recorded by a digital-to-analogue converter at a sampling rate of $64$ Hz, and the data are used in the analysis of plume dynamics. All measurements were conducted at a fixed input power at the bottom plate (or constant heat flux, 484 W), while maintaining the bulk temperature of the fluid at $40.0 ^o$C. The corresponding Rayleigh number was $Ra=6.18\times 10^9$ for the Newtonian case. To obtain sufficient statistics, each measurement lasted for 12 hours, corresponding to $\sim 1,400$ turnover times of the largest eddy in the system.

	Polyacrylimid (PAM; Polysciences 18522-100) with nominal molecular weight $Mw=18\times 10^6$ atomic mass unit (amu) was used in the experiment. The polymer concentration $c$ varied from 0 to 24 parts per million (ppm) by weight. The effects of polymer additives on the thermodynamic properties of water at such dilute concentrations are negligible except for the fluid viscosity \citep{Polymer_physics}. The measured fluid viscosity was increased by up to $30\%$ from the pure water value in the present study. 

	The Weissenberg number $Wi=\tau_p/\tau_{\eta}\sim 0.1$, where $\tau_p$ is the polymer relaxation time based on the Zimm model \citep{Zimm1956}, and $\tau_{\eta}$ is the Kolmogorov time scale directly measured using particle tracking velocimetry (PTV) at the cell centre. We remark that $Wi$ estimated above is a spatially- and temporally-averaged quantity. It is known that strong temperature and velocity fluctuations occur in turbulent RBC. Thus, instantaneous stretching of polymers can occur locally that cannot be reflected by this relatively small, spatial- and temporal-averaged $Wi$, which in fact is only a lower bound due to the polydispersity of the polymers \citep{2008NJP_Eberhard}. Additionally, we conducted experiments with a different kind of polymer, polyethylene oxide (PEO, $Mw=4\times10^6$ amu; Sigma Aldrich 189464), with $c$ varying from 0 ppm to 210 ppm. Because of the relatively small $Wi$ for PEO (0.02 in this case), the effects are not as pronounced as those with PAM and the results with PEO will only be discussed in the heat transport measurement.
	
	 As the fluid viscosity changed significantly when polymers were added, the Rayleigh number for each polymer concentration was calculated using the measured viscosity and temperature difference $\Delta T$. As a result, $Ra$ for PAM varied from $6.18\times 10^9$ to $7.43\times 10^9$ for $c$ between 0 ppm and 24 ppm, and for PEO it varied from $6.22\times10^9$ to $5.44\times10^9$ for $c$ between 0 ppm and 210 ppm. (Note that for both types of polymers, the viscosity $\nu$ and  temperature difference across plates $\Delta T$ both increase with the increase of $c$. In the case of PAM, the increase in $\Delta T$ more than compensates the increase in $\nu$ to result in an increase in $Ra$; whereas in the case of PEO, the combined results lead to a reduced $Ra$). In order to compare the measured heat transfers with those in the absence of polymers, local heat flux $\langle J_z(0)\rangle_t$ was measured as a function of $Ra$ without polymers and is used to normalize the measured heat flux $\langle J_z(c)\rangle_t$ at the same value of $Ra$. Here $\langle \cdots\rangle_t$ denotes time average.

\section{Results and discussions}\label{sec:Results_discussion}
\begin{figure}
	\begin{center}
	\includegraphics[width=\textwidth]{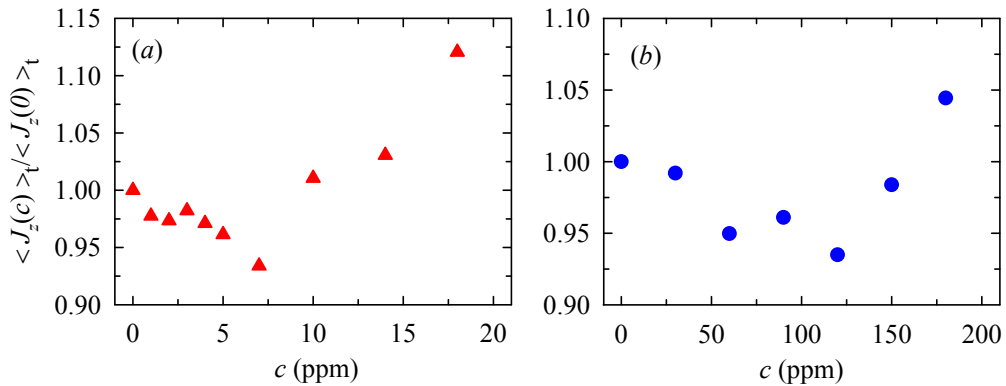}
    \caption{\label{figure:flux} Time-averaged vertical heat flux $\langle J_z(c)\rangle_t$ as a function of polymer concentration $c$, normalized by its Newtonian value $\langle J_z(0)\rangle_t$ measured at the same $Ra$. ($a$) PAM and ($b$) PEO.}
    	\end{center}
\end{figure}
	We first show in figure \ref{figure:flux} the time-averaged normalized vertical heat flux $\langle J_z(c)\rangle_t/\langle J_z(0)\rangle_t$ as a function of polymer concentration $c$  for PAM and PEO, respectively. It is seen that, with polymer additives, the local heat flux first decreases a little bit and then starts to increase above a certain polymer concentration. The maximum enhancement of the local $Nu$ is $\sim 12\%$ for PAM at $c=18$ ppm and $\sim 5\%$ for PEO at $c=180$ ppm. These observed local $Nu$ dependence on $c$ are qualitatively similar to the global heat transport measurement done in the same cell with PEO \citep{Wei2012PRE}. We note that if $c$ is further increased, i.e. beyond $18$ ppm for PAM and $180$ ppm for PEO, the heat transport drops.
 
\begin{figure}
	\begin{center}
		\includegraphics[width=\textwidth]{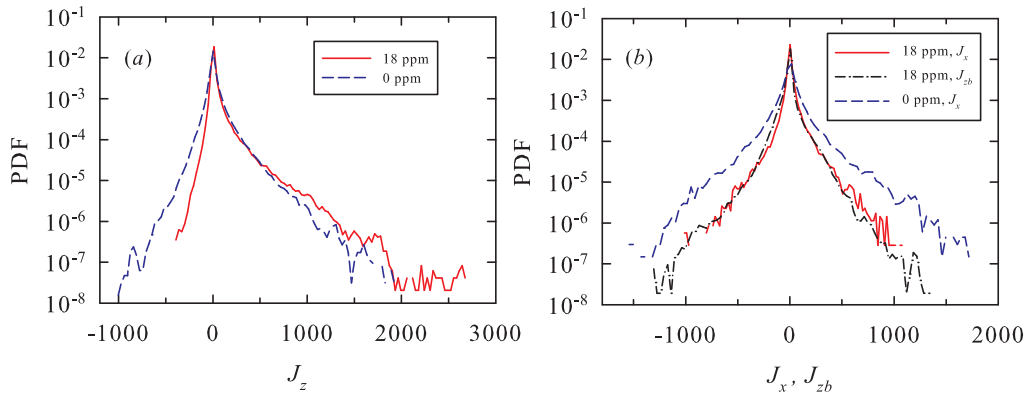}
		\caption{\label{figure:PDF_flux}($a$) Probability density functions (PDFs) of the vertical heat flux $J_z$ at $c=0$ and $c=18$ ppm; ($b$) PDFs of the horizontal heat flux $J_x$ at $c=0$ and $c=18$ ppm. The black dashed-dot line is the vertical background heat flux $J_{zb}$ at $c=18$ ppm. See text for definition of $J_{zb}$.}
\end{center}
\end{figure}

	To shed light on the mechanism of heat transport enhancement, we examine the probability density functions (PDFs) of the vertical heat flux $J_z$ and the horizontal one $J_x$. The PDFs of $J_z$ at $c=0$ and $18$ ppm are plotted in figure \ref{figure:PDF_flux}($a$). We first look at the Newtonian case. It is seen that the PDF is skewed toward positive values. The negative tail of the PDF arises from the random velocity and temperature fluctuations and is referred to as incoherent heat flux hereafter. The positive tail, on the other hand, contains contributions from both incoherent fluxes and coherent fluxes that arise from the positively correlated velocity and temperature signals. The difference between the positive and negative tails gives rise to the net heat transport \citep{Shang2003PRL,Shang2004PRE}. The figure shows that, with polymer additives, the negative tail is strongly suppressed on one hand and very large values ($\geq 500$) of instantaneous positive flux  become more probable on the other hand. Thus figure \ref{figure:PDF_flux}($a$) suggests that the effects of polymer additives are to suppress the incoherent heat transport, and to enhance the coherent heat transport at the same time.

	To confirm the above physical picture, we use a method introduced by \citet{Ching2004PRL} to decompose $J_z$ into heat flux carried by buoyant structures (i.e. plumes) and that by the turbulent background fluctuations, which is defined as 
	\begin{equation}
	J_{zb}(t)=(u_{zb}(t)\times \delta T(t))H/(\kappa\Delta T)
	\end{equation}
where $u_{zb}(t)=u_z(t)-\langle u_z|T(t)\rangle$ is the velocity related to the random background fluctuations. It has been found in the Newtonian case that $J_{zb}$ and $J_x$ collapse almost perfectly on top of each other, indicating that the incoherent fluxes arising from turbulent background fluctuations are isotropic \citep{Ching2004PRL}. A good collapse of $J_{zb}$ and $J_x$ is also observed in the present study for the case without polymers (not shown). 

	Figure \ref{figure:PDF_flux}($b$) shows the PDFs of $J_{zb}$ and $J_x$  measured at $c=18$ ppm, together with $J_x$ at $c=0$ ppm. It is first seen that, consistent with the behavior of the negative tail of the vertical heat flux shown in figure \ref{figure:PDF_flux}($a$), in the presence of polymers the horizontal heat transport arising from the random background fluctuations, e.g. the incoherent heat flux, is highly suppressed compared to that of the Newtonian case. In addition, the incoherent part of the vertical flux and the horizontal flux for the same polymer concentration essentially fall on top of each other, implying that in the presence of polymer additives heat flux arising from turbulent background fluctuations in the bulk of the cell remains isotropic as in the Newtonian case.

	As thermal plumes are the primary heat carriers in turbulent thermal convection \citep{Shang2003PRL}, the increase in the coherent part of the vertical flux suggests that polymer additives may have altered the properties of thermal plumes. To quantify this, we adopted a conditional averaging method similar to the one used by \citet{Huang2013PRL} to identify the hot (cold) plumes based on the physical intuition that the temperature of a hot (cold) plume must be higher (lower) than the surrounding fluid. Operationally, we define a thermal plume as a time period when $\pm[T(t)-\langle T\rangle_t]>b\sigma_T$ with $b=1.5$ being a threshold parameter, $\sigma_T=\sqrt{\langle(T-\langle T\rangle)^2\rangle}$ the root-mean-square (rms) temperature fluctuation, $``+"$ for hot plumes and $``-"$ for cold plumes. The width of a plume $\tau_{plume}$ is defined as the length of the consecutive time segment that satisfies the above criterion in the measured temperature time series. The amplitude of a plume $A_{plume}$ is then defined as the absolute value of the largest temperature deviation from the mean within this time segment. We have tried different $b$ from 0.8 to 3 and the statistical properties based on different $b$ remain unchanged.

\begin{figure}
\begin{center}
		\includegraphics[width=\textwidth]{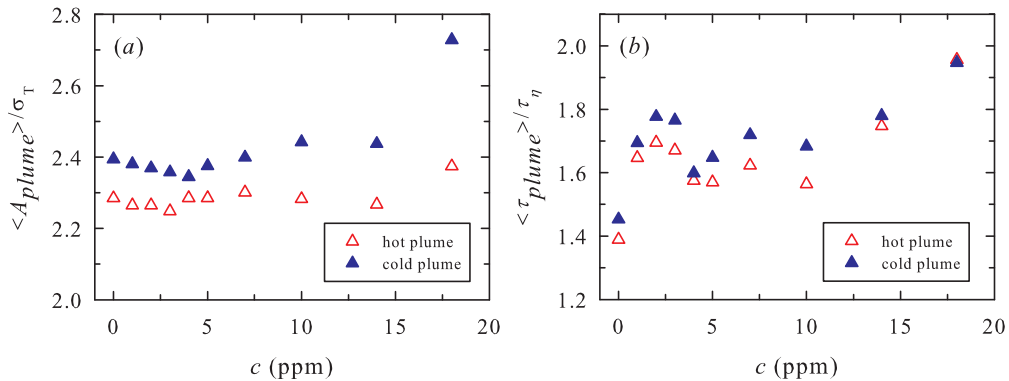}		
		\caption{\label{figure:plume_a_tau} The average normalized plume amplitude $\langle A_{plume}\rangle/\sigma_T$ ($a$) and plume width $\langle\tau_{plume}\rangle/\tau_{\eta}$ ($b$)  as a function of polymer concentration $c$, where $\sigma_T$ and $\tau_{\eta}$ are, respectively, the root-mean-square temperature fluctuation and the Kolmogorov time scale.}
\end{center}
\end{figure}

\begin{figure}
	\begin{center}
		\includegraphics[width=\textwidth]{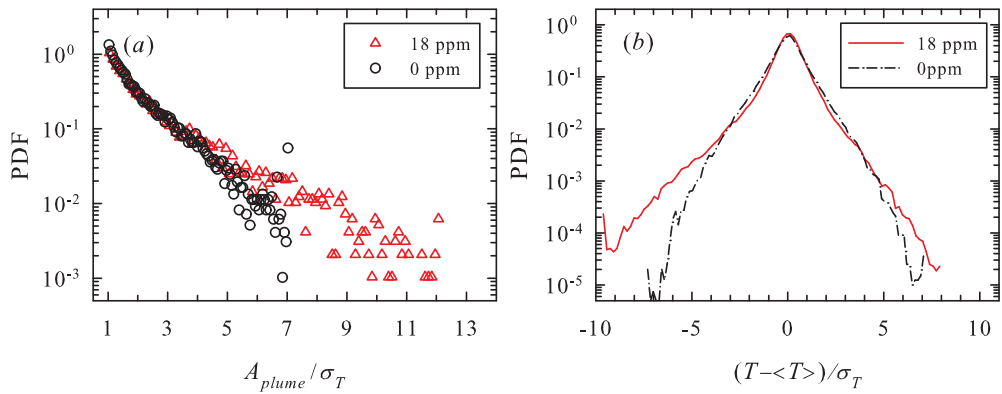}
		\caption{\label{figure:PDF_a_T}Probability density functions (PDFs) of ($a$) the normalized plume amplitude $A_{plume}/\sigma_T$ and of ($b$) the normalized temperature fluctuation $(T-\langle T\rangle)/\sigma_T$.}
\end{center}
\end{figure}

	In figure \ref{figure:plume_a_tau}($a$) we plot the normalized average plume amplitude $\langle A_{plume}\rangle/\sigma_T$ as a function of $c$. It is seen that the dependence of plume amplitude $ A_{plume}$ on $c$ has a qualitatively similar trend as that of local $Nu$, i.e. the plume amplitude first drops a bit when polymers are added and then increases when $c$ is beyond a certain value. Similar trend has been observed for the average normalized plume width $\langle\tau_{plume}\rangle/\tau_{\eta}$ as shown in figure \ref{figure:plume_a_tau}($b$), i.e. with the increase of $c$, $\tau_{plume}$ increases. The PDFs of $A_{plume}/\sigma_T$ for $c=0$ and 18 ppm are shown in figure \ref{figure:PDF_a_T}($a$). It is seen that with polymer additives plumes with larger amplitudes become more probable compared to the Newtonian case. The increased $A_{plume}$ implies that large temperature deviations from the mean would be more probable with polymer additives, which is confirmed by the higher tails of the PDF of the normalized temperature fluctuations for $c=18$ ppm compared to the Newtonian case as shown in figure \ref{figure:PDF_a_T}($b$). A feature we have noticed is that the increased temperature deviation from the mean or the increase in plume amplitude only becomes significant when $c$ is sufficiently high. For example, at $c=7$ ppm the temperature PDF is essentially the same as in the $c=0$ case (not shown). Both larger $A_{plume}$ and $\tau_{plume}$ imply that thermal plumes have become more coherent with polymer additives. 
	
\begin{figure}
	\begin{center}
		\includegraphics[width=\textwidth]{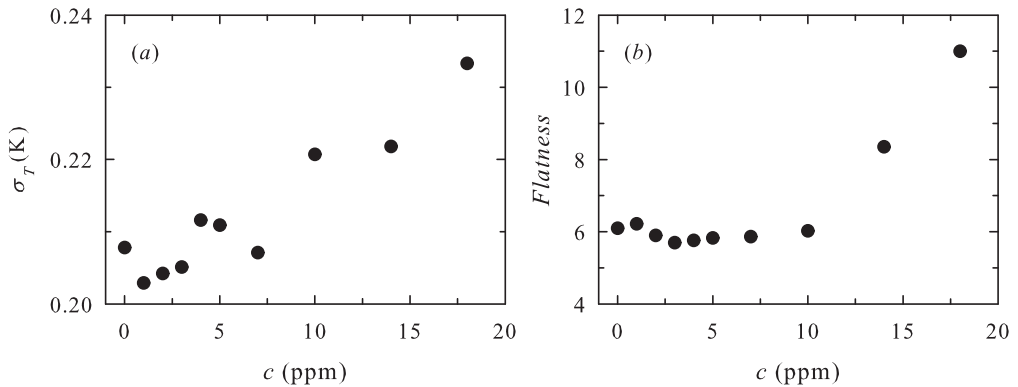}
		\caption{\label{figure:SigmaT} The temperature rms $\sigma_T$ ($a$) and \textit{Flatness} ($b$) as a function of polymer concentration $c$.}
\end{center}
\end{figure}  
	
	One may also notice from figure \ref{figure:PDF_a_T}($b$) that the PDF at $c=18$ ppm is asymmetric, indicating that the cold plumes undergo more significant changes than the hot ones. This different behaviour is also revealed $A_{plume}$ and $\tau_{plume}$ (see figure \ref{figure:plume_a_tau}). Recently, \citet{Traore2015} have shown that $\tau_p$ of PAM has a very strong negative dependence on temperature. For example, they have found that $\tau_p$ of PAM can increase by two orders of magnitude when temperature is decreased from $\sim 35^{o}$C to $\sim 5^{o}$C. Therefore, polymers in a parcel of hot fluid will interact with the turbulent flow differently than those in a parcel of cold fluid. This also supports the idea of a local $Wi$ that can be much larger than the one based on globally averaged properties. 
	
	We also found that the local temperature fluctuation $\sigma_T$ increases with $c$ as shown in figure \ref{figure:SigmaT}($a$), and has similar $c$ dependence as $J_z$. Similar observation was reported in a DNS study of homogeneous RBC with polymer additives \citep{Benzi2010PRL}. Moreover, the temperature flatness, which is defined as $\langle( T-\langle T\rangle)^4 \rangle/\sigma_T^4$, shown in figure \ref{figure:SigmaT}($b$), increases from 6 at $c=0$ ppm to 11 at $c=18$ ppm, also suggesting that the rare but large deviations from the mean are more probable with polymer additives.

\begin{figure}
\begin{center}
		\includegraphics[width=\textwidth]{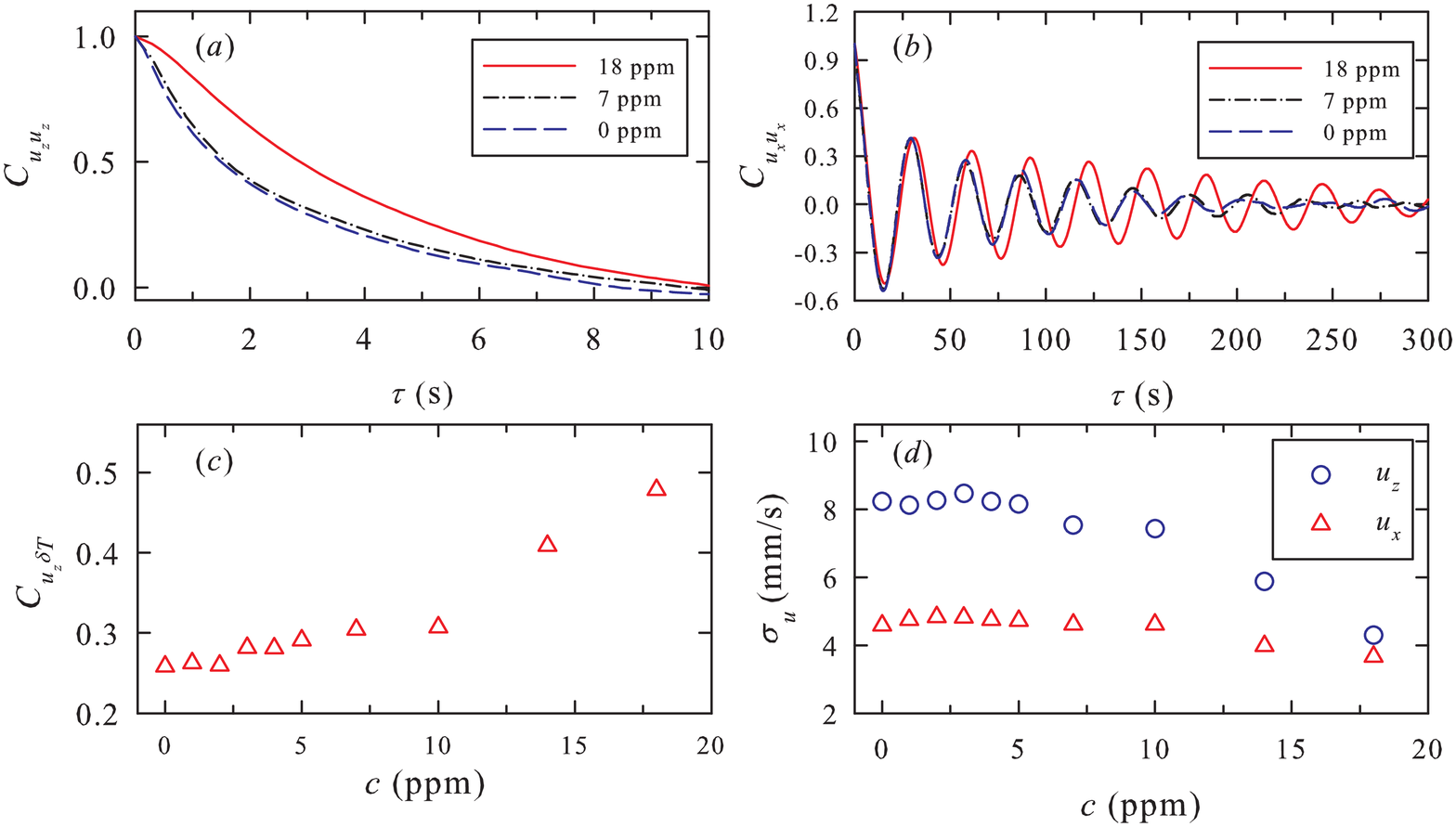}
		\caption{\label{figure:corr_velocity} Auto-correlation functions of the vertical velocity $u_z$ ($a$) and the horizontal velocity $u_x$ ($b$) at polymer concentration $c=0$, $7$ and $18$ ppm. ($c$) Correlation coefficient between temperature deviation $\delta T$ from the mean and the vertical velocity $u_z$ as a function of $c$. ($d$) The rms velocity $\sigma_u$ as a function of $c$ for the vertical and horizontal velocities, respectively.}
\end{center}
\end{figure} 

	In addition to the temperature field, the velocity field also becomes more coherent with polymer additives, which is quantified by the auto-correlation functions of the vertical velocity $u_z$ and the horizontal one $u_x$ measured at different $c$. It is seen from figure \ref{figure:corr_velocity}($a$) that with increasing $c$, the decay of the auto-correlation functions for $u_z$ becomes slower, suggesting an enhancement of the coherency of velocity field. The enhanced coherence in $u_x$ is manifested by the increased number of oscillation periods in the auto-correlation function shown in figure \ref{figure:corr_velocity}($b$). Figure \ref{figure:corr_velocity}($c$) shows that the degree of correlation between the velocity and temperature increases with increasing polymer concentration, which is another evidence that the coherence of thermal plumes is increased by polymer additives, as the velocity and temperature of thermal plumes should be correlated. Again, we note from the above that the increase of plume coherency only becomes significant when $c$ is above a certain threshold value. At lower $c$, these properties remain essentially the same as in the Newtonian case. In figure \ref{figure:corr_velocity}($d$) we show the rms velocity $\sigma_u$ as a function of $c$. It is seen that $\sigma_u$ decreases with increasing $c$, suggesting that the velocity fluctuations are suppressed by polymers.
	
	\begin{figure}
       \begin{center}
		\includegraphics[width=\textwidth]{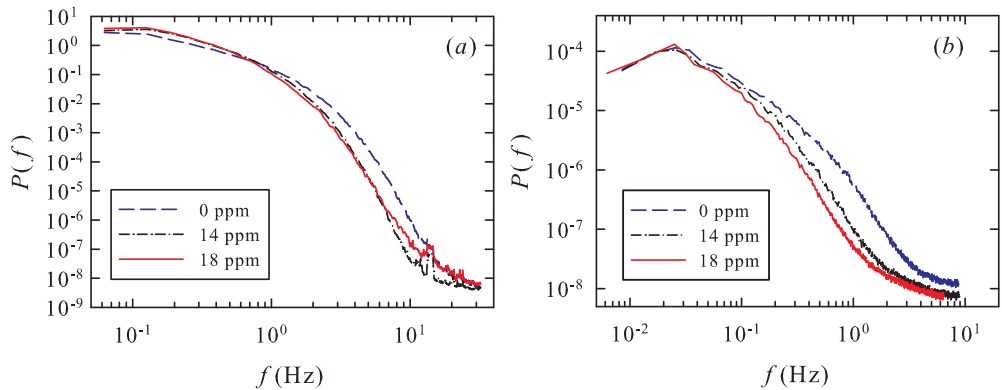}
		\caption{\label{figure:spectra}($a$) Power spectra of the temperature at $c=0$, $14$ and $18$ ppm. ($b$) The corresponding power spectra of the velocity.}
       \end{center}
\end{figure}

	To gain further insight into the possible mechanism that leads to the enhancement of plume coherency, we examine the power spectra of both the temperature and velocity fluctuations for several polymer concentrations as shown in figure \ref{figure:spectra}. The magnitudes of both temperature and velocity spectra at high frequencies are seen to decrease significantly with increasing $c$, which implies that polymers suppress turbulent fluctuations at small scales. Similar observations were reported in a DNS study of Rayleigh-Taylor turbulence \citep{Boffetta2010PRL}. Note that there is a crossover for the temperature power spectra for cases with and without polymer at $f_c \simeq 0.7$ Hz. For frequencies smaller than $f_c$ (corresponding to larger length scales), the energy contained in these scales is larger than the Newtonian case and it is smaller for frequencies larger than $f_c$, suggesting that there is a redistribution of thermal energy among different scales when polymers are added to the flow. For the velocity power spectra, a suppression of turbulent kinetic energy at small scales is also observed. However, the situation is somewhat different, in the sense that a crossover to energy at the large scale is not observed. It is seen that the turbulent kinetic energy decreases at all scales smaller than the energy injection scale when polymers are added, and the degree of energy quenching increases with increase of $c$. The different behaviours of the temperature and velocity power spectra at small frequencies (large scales) are consistent with the observation in \citet{Benzi2010PRL}, where it is found that polymers strongly modify the large scale of temperature field, while having negligible effects on the large scale of velocity field. There have been evidences that the energy of turbulent fluctuations in turbulent RBC is supplied by thermal plumes \citep{Xia2003PRE}. When polymers are added to the flow, the small-scale turbulent fluctuations are suppressed as suggested by the reduction of velocity fluctuation $\sigma_u$ shown in figure \ref{figure:corr_velocity}($d$) and the power spectra of velocity fluctuations in figure \ref{figure:spectra}($b$). Thermal plumes thus supply less energy to the turbulence, and they are more energetic and able to transport heat more efficiently.

\begin{figure}
\begin{center}
		\includegraphics[width=0.6\textwidth]{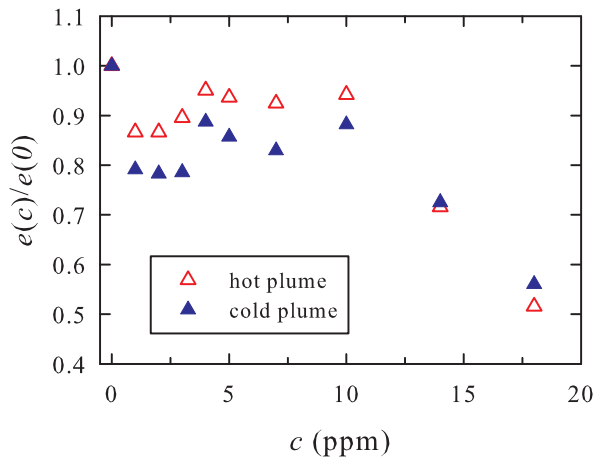}
		\caption{\label{figure:plume_emission} Plume emission rate $e(c)$ normalized by the Newtonian value $e(0)$ as a function of polymer concentration $c$.}
\end{center}
\end{figure}

We note that as the plume emissions are manifestations of thermal boundary layer instability in turbulent RBC \citep{Malkus}, the effects of polymer additives on boundary layers may be examined indirectly by estimating the plume emission rate $e$. Using the extracted plumes aforementioned we obtain plume emission rate $e$ in unit of plume number per hour. The dependence on polymer concentration $c$ of $e(c)$ normalized by its Newtonian value $e(0)$ is shown in figure \ref{figure:plume_emission}. It is clearly seen that with the increase of $c$, $e$ decreases. Remarkably, $e$ is very sensitive to polymer additives, e.g. upon the addition of 1 ppm of polymers, $e$ drops by $\sim 15\%$ comparing to the Newtonian case. These observations provide an evidence for the proposed stabilization of thermal boundary layer and the resulting reduction of plume emissions by polymer additives \citep{Ahlers2010PRL,Wei2012PRE}. 

Finally, we remark that there appears to exist a certain polymer concentration, only beyond which can significant heat transport enhancement be achieved. The existence of such a threshold concentration may be seen from a number of figures, such as figure \ref{figure:flux} where it is seen to be around 7 ppm for PAM and around 120 ppm for PEO. This threshold $c$, though not very sharp, may be understood in such a way that the enhancement of plume coherency requires more polymers than the stabilization of the boundary layer, which is supported by the $c-$dependence of $A_{plume}$ and $\tau_{plume}$ (figure \ref{figure:plume_a_tau}) as well as temperature rms $\sigma_T$ (figure \ref{figure:SigmaT}) and the auto-correlation function of the velocity $u_{\alpha}$ (figure \ref{figure:corr_velocity}($a$, $b$)), all of which undergo significant changes only when $c$ is beyond a certain value. Therefore, the local heat flux first decreases gradually due to the reduction of plume emissions before the significant enhancement starts.  

\section{Concluding remarks}
	In summary, we have demonstrated that polymer additives can significantly enhance heat transport in the bulk of turbulent thermal convection. The enhancement occurs concurrently with the increased coherency of thermal plumes. It is found that the effects of polymer additives in the bulk of turbulent thermal convection are to \textit{suppress} the random turbulent fluctuations on one hand, and to \textit{increase} the coherence of the temperature and velocity fields on the other hand, i.e. it \textit{suppresses} the incoherent heat transfer and \textit{enhances} the coherent one. The effects of polymer additives on the boundary layer are indirectly examined by studying the plume emission rate $e$. The reduction of $e$ with the increase of $c$ indicates a stabilization of the thermal boundary layer by polymer additives. This can be achieved through two routes. One is through the reduced bulk velocity fluctuations that constantly perturb the BL \citep{Zhou2010PRL}; the other is through the
increased viscosity, which leads to a more stable viscous BL. As the viscous and thermal
boundary layers are strongly coupled \citep{Zhou2010JFM}, this will result in a more stable thermal BL. In this context, the present experiment supports the idea that the combined effects of enhancement of plume coherency in the turbulent bulk flow and reduction of plume emission in the boundary layer determine whether the global heat transport will be enhanced or reduced in turbulent thermal convection with polymer additive \citep{Ahlers2010PRL,Benzi2010PRL,Wei2012PRE}. Finally, we remark that the decrease of $J_z$ beyond $c=18$ ppm for PAM and 180 ppm for PEO may be viewed as the result that the effect of reduced plume emission is overtaking the effect of enhanced plume coherency.
	
\section*{Acknowledgement}
	This work was supported in part by the Hong Kong Research Grant Council under Grant Nos. CUHK 403712 and 404513; and in part by a grant from the PROCORE-France/Hong Kong Joint Research Scheme sponsored by the Research Grants Council of Hong Kong and the Consulate General of France in Hong Kong (Reference No. F-HK19/11T)

\bibliographystyle{jfm}

\begin{thebibliography}{14}
\expandafter\ifx\csname natexlab\endcsname\relax\def\natexlab#1{#1}\fi

\bibitem[Batchelor(1971)]{Batchelor59}
{\sc Batchelor, G.~K.} 1971 Small-scale variation of convected quantities like
  temperature in turbulent fluid. part 1. general discussion and the case of
  small conductivity. {\em J.~Fluid Mech.\/} {\bf 5}, 113--133.

\bibitem[Brownell \& Su(2004)]{Brownell04}
{\sc Brownell, C.~J. \& Su, L.~K.} 2004 Planar measurements of differential
  diffusion in turbulent jets. {\em AIAA Paper 2004-2335\/}.

\bibitem[Brownell \& Su(2007)]{Brownell07}
{\sc Brownell, C.~J. \& Su, L.~K.} 2007 Scale relations and spatial spectra in
  a differentially diffusing jet. {\em AIAA Paper 2007-1314\/}.

\bibitem[Dennis(1985)]{Dennis85}
{\sc Dennis, S. C.~R.} 1985 {Compact explicit finite difference approximations
  to the Navier--Stokes equation}. In {\em Ninth Intl Conf. on Numerical
  Methods in Fluid Dynamics\/} (ed. Soubbaramayer \& J.~P. Boujot), {\em
  Lecture Notes in Physics\/}, vol. 218, pp. 23--51. Springer.

\bibitem[Hwang \& Tuck(1970)]{Hwang70}
{\sc Hwang, L.-S. \& Tuck, E.~O.} 1970 On the oscillations of harbours of
  arbitrary shape. {\em J.~Fluid Mech.\/} {\bf 42}, 447--464.

\bibitem[Koch(1983)]{Koch83}
{\sc Koch, W.} 1983 Resonant acoustic frequencies of flat plate cascades. {\em
  J.~Sound Vib.\/} {\bf 88}, 233--242.

\bibitem[Lee(1971)]{Lee71}
{\sc Lee, J.-J.} 1971 Wave-induced oscillations in harbours of arbitrary
  geometry. {\em J.~Fluid Mech.\/} {\bf 45}, 375--394.

\bibitem[Linton \& Evans(1992)]{Linton92}
{\sc Linton, C.~M. \& Evans, D.~V.} 1992 The radiation and scattering of
  surface waves by a vertical circular cylinder in a channel. {\em Phil.\
  Trans.\ R. Soc.\ Lond.\/} {\bf 338}, 325--357.

\bibitem[Martin(1980)]{Martin80}
{\sc Martin, P.~A.} 1980 On the null-field equations for the exterior problems
  of acoustics. {\em Q.~J. Mech.\ Appl.\ Maths\/} {\bf 33}, 385--396.

\bibitem[Miller(1991)]{Miller91}
{\sc Miller, P.~L.} 1991 Mixing in high schmidt number turbulent jets. PhD
  thesis, California Institute of Technology.

\bibitem[Rogallo(1981)]{Rogallo81}
{\sc Rogallo, R.~S.} 1981 Numerical experiments in homogeneous turbulence. {\em
  Tech. Rep.\/} 81835. NASA Tech.\ Mem.

\bibitem[Ursell(1950)]{Ursell50}
{\sc Ursell, F.} 1950 Surface waves on deep water in the presence of a
  submerged cylinder i. {\em Proc.\ Camb.\ Phil.\ Soc.\/} {\bf 46}, 141--152.

\bibitem[{van Wijngaarden}(1968)]{Wijngaarden68}
{\sc {van Wijngaarden}, L.} 1968 On the oscillations near and at resonance in
  open pipes. {\em J.~Engng Maths\/} {\bf 2}, 225--240.

\bibitem[Worster(1992)]{Worster92}
{\sc Worster, M.~G.} 1992 {The dynamics of mushy layers}. In {\em In
  Interactive dynamics of convection and solidification\/} (ed. S.~H. Davis,
  H.~E. Huppert, W.~Muller \& M.~G. Worster), pp. 113--138. Kluwer.

\end{thebibliography}


\begin{thebibliography}{29}
\expandafter\ifx\csname natexlab\endcsname\relax\def\natexlab#1{#1}\fi
\def\au#1{#1} \def\ed#1{#1} \def\yr#1{#1}\def\at#1{#1}\def\jt#1{\textit{#1}}
  \def\bt#1{#1}\def\bvol#1{\textbf{#1}} \def\vol#1{#1} \def\pg#1{#1}
  \def\publ#1{#1}\def\arxiv#1{#1}\def\org#1{#1}\def\st#1{\textit{#1}}

\bibitem[Ahlers {\em et~al.\/}(2009)Ahlers, Grossmann \& Lohse]{ahlers2009RMP}
{\sc \au{Ahlers, G.}, \au{Grossmann, S.} \& \au{Lohse, D.}} \yr{2009}  \at{Heat
  transfer and large-scale dynamics in turbulent {R}ayleigh-{B}\'enard
  convection}.  \jt{Rev. Mod. Phys.}  \bvol{81},  \pg{503--537}.

\bibitem[Ahlers \& Nikolaenko(2010)]{Ahlers2010PRL}
{\sc \au{Ahlers, G.} \& \au{Nikolaenko, A.}} \yr{2010}  \at{Effect of a polymer
  additive on heat transport in turbulent {R}ayleigh-{B}\'enard convection}.
  \jt{Phys. Rev. Lett.}  \bvol{104},  \pg{034503}.

\bibitem[Benzi {\em et~al.\/}(2010)Benzi, Ching \& De~Angelis]{Benzi2010PRL}
{\sc \au{Benzi, R.}, \au{Ching, E. S.~C.} \& \au{De~Angelis, E.}} \yr{2010}
  \at{Effect of polymer additives on heat transport in turbulent thermal
  convection}.  \jt{Phys. Rev. Lett.}  \bvol{104},  \pg{024502}.

\bibitem[Boffetta {\em et~al.\/}(2010)Boffetta \textit{et~al.}]{Boffetta2010PRL}
{\sc \au{Boffetta, G.}, \au{Mazzino, A.}, \au{Musacchio, S.} \& \au{Vozella,
  L.}} \yr{2010}  \at{Polymer heat transport enhancement in thermal convection:
  The case of {R}ayleigh-{T}aylor turbulence}.  \jt{Phys. Rev. Lett.}
  \bvol{104},  \pg{184501}.

\bibitem[Calzavarini {\em et~al.\/}(2005)Calzavarini \em et~al.]{2005PoF_Cal}
{\sc \au{Calzavarini, E.}, \au{Lohse, D.}, \au{Toschi, F.} \& \au{Tripiccione,
  R.}} \yr{2005}  \at{{R}ayleigh and {P}randtl number scaling in the bulk of
  {R}ayleigh-{B}\'enard turbulence}.  \jt{Phys. Fluids}  \bvol{17}~(5).

\bibitem[Chill\'a \& Schumacher(2012)]{SchumacherChilla2012}
{\sc \au{Chill\'a, F.} \& \au{Schumacher, J.}} \yr{2012}  \at{New perspectives
  in turbulent {R}ayleigh-{B}\'enard convection}.  \jt{Euro. Phys. J. E}
  \bvol{35}~(7),  \pg{1--25}.

\bibitem[Ching {\em et~al.\/}(2004)Ching, Guo, Shang, Tong \&
  Xia]{Ching2004PRL}
{\sc \au{Ching, E. S.~C.}, \au{Guo, H.}, \au{Shang, X.-D.}, \au{Tong, P.} \&
  \au{Xia, K.-Q.}} \yr{2004}  \at{Extraction of plumes in turbulent thermal
  convection}.  \jt{Phys. Rev. Lett.}  \bvol{93},  \pg{124501}.

\bibitem[Crawford {\em et~al.\/}(2008)Crawford \em et al.]{2008NJP_Eberhard}
{\sc \au{Crawford, A.~M.}, \au{Mordant, N.}, \au{Xu, H.} \& \au{Bodenschatz,
  E.}} \yr{2008}  \at{Fluid acceleration in the bulk of turbulent dilute
  polymer solutions}.  \jt{New J. Phys.}  \bvol{10}~(12),  \pg{123015}.

\bibitem[Huang {\em et~al.\/}(2013)Huang, Kaczorowski, Ni \& Xia]{Huang2013PRL}
{\sc \au{Huang, S.-D.}, \au{Kaczorowski, M.}, \au{Ni, R.} \& \au{Xia, K.-Q.}}
  \yr{2013}  \at{Confinement-induced heat-transport enhancement in turbulent
  thermal convection}.  \jt{Phys. Rev. Lett.}  \bvol{111},  \pg{104501}.

\bibitem[Lohse \& Toschi(2003)]{2003PRL_Lohse}
{\sc \au{Lohse, D.} \& \au{Toschi, F.}} \yr{2003}  \at{Ultimate state
  of thermal convection}.  \jt{Phys. Rev. Lett.}  \bvol{90},  \pg{034502}.

\bibitem[Lohse \& Xia(2010)]{Xia2010ARFM}
{\sc \au{Lohse, D.} \& \au{Xia, K.-Q.}} \yr{2010}  \at{Small-scale properties
  of turbulent {R}ayleigh-{B}\'enard convection}.  \jt{Annu. Rev. Fluid Mech.}
  \bvol{42}~(1),  \pg{335--364}.

\bibitem[Malkus(1954)]{Malkus}
{\sc \au{Malkus, W. V.~R.}} \yr{1954}  \at{The heat transport and spectrum of
  thermal turbulence}.  \jt{Proceedings of the Royal Society of London A:
  Mathematical, Physical and Engineering Sciences}  \bvol{225}~(1161),
  \pg{196--212}.

\bibitem[Ouellette {\em et~al.\/}(2009)Ouellette, Xu \&
  Bodenschatz]{2007_JFM_Ouelltte}
{\sc \au{Ouellette, N.~T.}, \au{Xu, H.} \& \au{Bodenschatz, E.}} \yr{2009}
  \at{Bulk turbulence in dilute polymer solutions}.  \jt{J. Fluid Mech.}
  \bvol{629},  \pg{375--385}.

\bibitem[Procaccia {\em et~al.\/}(2008)Procaccia, L'vov \&
  Benzi]{Procaccia2008RMP}
{\sc \au{Procaccia, I.}, \au{L'vov, V.~S.} \& \au{Benzi, R.}} \yr{2008}
  \at{Theory of drag reduction by polymers in wall-bounded turbulence}.
  \jt{Rev. Mod. Phys.}  \bvol{80},  \pg{225--247}.

\bibitem[Qiu {\em et~al.\/}(2005)Qiu, Xia \& Tong]{Qiu2005JoT}
{\sc \au{Qiu, X.-L.}, \au{Xia, K.-Q.} \& \au{Tong, P.}} \yr{2005}
  \at{Experimental study of velocity boundary layer near a rough conducting
  surface in turbulent natural convection}.  \jt{J. Turb.}  \bvol{6},
  \pg{N30}.

\bibitem[Rubinstein \& Colby(2003)]{Polymer_physics}
{\sc \au{Rubinstein, M.} \& \au{Colby, R.~H.}} \yr{2003} {\em Polymer
  Physics\/}.  \publ{Oxford University Press, Oxford, UK}.

\bibitem[Shang {\em et~al.\/}(2003)Shang, Qiu, Tong \& Xia]{Shang2003PRL}
{\sc \au{Shang, X.-D.}, \au{Qiu, X.-L.}, \au{Tong, P.} \& \au{Xia, K.-Q.}}
  \yr{2003}  \at{Measured local heat transport in turbulent
  {R}ayleigh-{B}\'enard convection}.  \jt{Phys. Rev. Lett.}  \bvol{90},
  \pg{074501}.

\bibitem[Shang {\em et~al.\/}(2004)Shang, Qiu, Tong \& Xia]{Shang2004PRE}
{\sc \au{Shang, X.-D.}, \au{Qiu, X.-L.}, \au{Tong, P.} \& \au{Xia, K.-Q.}}
  \yr{2004}  \at{Measurements of the local convective heat flux in turbulent
  {R}ayleigh-{B}\'enard convection}.  \jt{Phys. Rev. E}  \bvol{70},
  \pg{026308}.

\bibitem[Sun {\em et~al.\/}(2008)Sun, Cheung \& Xia]{Sun2008JFM}
{\sc \au{Sun, C.}, \au{Cheung, Y.-H.} \& \au{Xia, K.-Q.}} \yr{2008}
  \at{Experimental studies of the viscous boundary layer properties in
  turbulent {R}ayleigh-{B}\'enard convection}.  \jt{J. Fluid Mech.}
  \bvol{605},  \pg{79--113}.

\bibitem[Toms(1948)]{toms1948some}
{\sc \au{Toms, B.}} \yr{1948} Some observations on the flow of linear polymer
  solutions through straight tubes at large {R}eynolds numbers.  \bt{In {\em
  Proc. Int. Congr. Rheo.\/}},  \pg{pp. 135--141}.

\bibitem[Traore {\em et~al.\/}(2015)Traore, Castelain \& Burghelea]{Traore2015}
{\sc \au{Traore, B.}, \au{Castelain, C.} \& \au{Burghelea, T.}} \yr{2015}
  \at{Efficient heat transfer in a regime of elastic turbulence}.  \jt{J.
  Non-Newtonian Fluid Mech.}  \bvol{223},  \pg{62 -- 76}.

\bibitem[Wei {\em et~al.\/}(2014)Wei, Chan, Ni, Zhao \& Xia]{Wei2014JFM}
{\sc \au{Wei, P.}, \au{Chan, T.-S.}, \au{Ni, R.}, \au{Zhao, X.-Z.} \& \au{Xia,
  K.-Q.}} \yr{2014}  \at{Heat transport properties of plates with smooth and
  rough surfaces in turbulent thermal convection}.  \jt{J. Fluid Mech.}
  \bvol{740},  \pg{28--46}.

\bibitem[Wei {\em et~al.\/}(2012)Wei, Ni \& Xia]{Wei2012PRE}
{\sc \au{Wei, P.}, \au{Ni, R.} \& \au{Xia, K.-Q.}} \yr{2012}  \at{Enhanced and
  reduced heat transport in turbulent thermal convection with polymer
  additives}.  \jt{Phys. Rev. E}  \bvol{86},  \pg{016325}.

\bibitem[Xi {\em et~al.\/}(2013)Xi, Bodenschatz \& Xu]{2013_PRL_Xi}
{\sc \au{Xi, H.-D.}, \au{Bodenschatz, E.} \& \au{Xu, H.}} \yr{2013}
  \at{Elastic energy flux by flexible polymers in fluid turbulence}.  \jt{Phys.
  Rev. Lett.}  \bvol{111},  \pg{024501}.

\bibitem[Xia(2013)]{Xia2013TAML}
{\sc \au{Xia, K.-Q.}} \yr{2013}  \at{Current trends and future directions in
  turbulent thermal convection}.  \jt{Theor. Appl. Mech. Lett.}  \bvol{3},
  \pg{052001}.

\bibitem[Xia {\em et~al.\/}(2003)Xia, Sun \& Zhou]{Xia2003PRE}
{\sc \au{Xia, K.-Q.}, \au{Sun, C.} \& \au{Zhou, S.-Q.}} \yr{2003}  \at{Particle
  image velocimetry measurement of the velocity field in turbulent thermal
  convection}.  \jt{Phys. Rev. E}  \bvol{68},  \pg{066303}.

\bibitem[Zhou {\em et~al.\/}(2010)Zhou \em et~al.]{Zhou2010JFM}
{\sc \au{Zhou, Q.}, \au{Stevens, R. J. A.~M.}, \au{Sugiyama, K.},
  \au{Grossmann, S.}, \au{Lohse, D.} \& \au{Xia, K.-Q.}} \yr{2010}
  \at{Prandtl-{B}lasius temperature and velocity boundary-layer profiles in
  turbulent {R}ayleigh-{B}\'enard convection}.  \jt{J. Fluid Mech.}
  \bvol{664},  \pg{297--312}.

\bibitem[Zhou \& Xia(2010)]{Zhou2010PRL}
{\sc \au{Zhou, Q.} \& \au{Xia, K.-Q.}} \yr{2010}  \at{Measured instantaneous
  viscous boundary layer in turbulent {R}ayleigh-{B}\'enard convection}.
  \jt{Phys. Rev. Lett.}  \bvol{104},  \pg{104301}.

\bibitem[Zimm(1956)]{Zimm1956}
{\sc \au{Zimm, Bruno~H.}} \yr{1956}  \at{Dynamics of polymer molecules in
  dilute solution: Viscoelasticity, flow birefringence and dielectric loss}.
  \jt{J. Chem. Phys.}  \bvol{24},  \pg{269--278}.

\end{thebibliography}

\end{document}